# Optimization and numerical simulation for the accelerator of the commercial H⁻ cyclotron ion source

M.Hosseinzadeh[1;1)], H.Afarideh[1;2)]

[1]Nuclear Engineering and Physics Department, Amirkabir University of Technology, Tehran, 15875-4413, Iran

**Abstract:** A new ion source will be prepared for the CYCLONE30 commercial cyclotron with a much advanced performance compared with the previous one. The newly designed ion source has a very large transparency without deteriorating the beam optics, which is designed to deliver an H- beam at 30 keV. The plasma generator of the ion source is of an axially cusped bucket type, and the whole inner wall, except the cathode filaments and plasma electrode side, functions as an anode. The accelerator assembly consists of three circular aperture electrodes made of copper. The simulation study was focused on finding parameter sets that raise the percent of beam transmitted as large as possible and reduces the beam divergence as low as possible. From the simulation results, it was concluded that it is possible to achieve this goal by sliming the plasma electrode (G1), shortening the first gap (G1-G2), and adjusting the G2 voltage.

**Key words:** multicusp ion source, extraction, electron tracking, perveance, angular divergence.

## 1 Introduction

The CYCLONE30 commercial accelerator utilizes an ion source technology similar to a LBL style ion source dating from the early 1980s [1]. The 30 keV H- beam current produced by this ion source is typically about 2.5 mA with an emittance of 400 mm mrad [2]. In order to reduce angular divergence and increase the transported beam current from the CYCLONE30 multicusp-type H− ion source, improvements to the extraction system are required. To facilitate this, the present geometry requires optimization. In this paper, simulation studies to find the optimized design features of the new ion source and the possible electrode geometry modifications needed to extract the highest quality beam are described. The accelerator upgrade concept was determined theoretically by simulations using the CST PS code.

## 2 Description of plasma generator and present extraction system

The CYCLONE30 ion source is, mechanically and functionally, composed of two main parts: the plasma generator and the ion beam accelerator. The plasma generator is a 150 mm copper cylinder, 100 mm in diameter around which ten columns of permanent magnets with magnetic field of 6.2 kG are mounted [3].

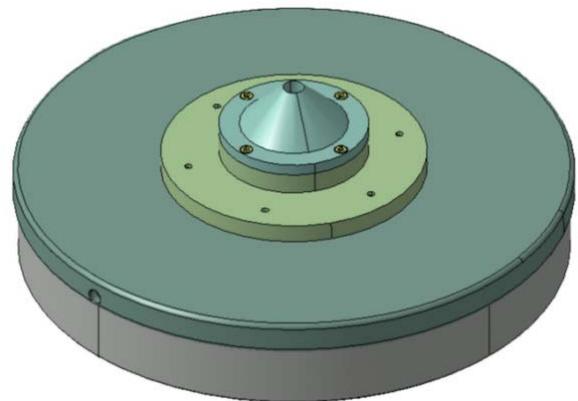

Fig.1. Schematic layout of the CYCLONE30 accelerator system.

Cusp plasma confinement is achieved in the usual manner with ten rows of permanent magnets around the circumference of the chamber, and two lines of cusp magnets in the back plate. A tungsten filament mounted to two copper posts that extend a significant distance into the source volume is used. The ion beam accelerator has a triode electrode column. The dimensions of the present accelerating column are given in Fig. 2. The first electrode or plasma electrode, G1, defines the boundary of the source plasma, from which ions are extracted. The first gap extracts ions through G1 apertures and the second gap post accelerates the ion beam to a desired level. Second and third electrodes in Fig.2. are identified as G2 and G3 in the rest of the paper. As is shown in Fig.2. the plasma electrode thickness is 10mm, the distance between first electrode and second electrode, D12, is 3.5mm and the voltage applied to the second electrode is -26000V. Also the plasma electrode inclination angle which affects beam divergence is 30°. Fig.3. illustrates electrical potential lines of these three electrodes plotted by CST PS program.

---
1) E-mail: M.Hosseinzadeh5176@gmail.com
2) E-mail: Hafarideh@aut.ac.ir

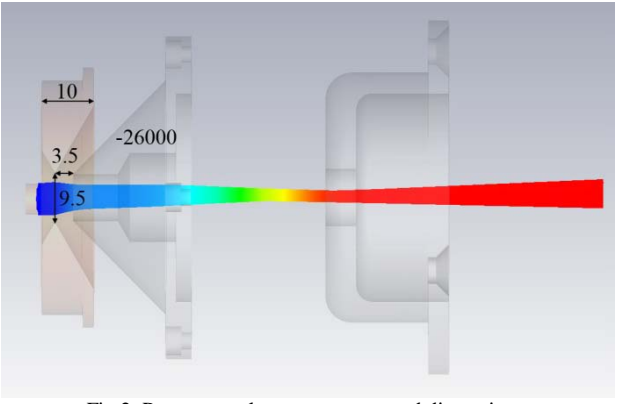

Fig.2. Present accelerator structure and dimensions.

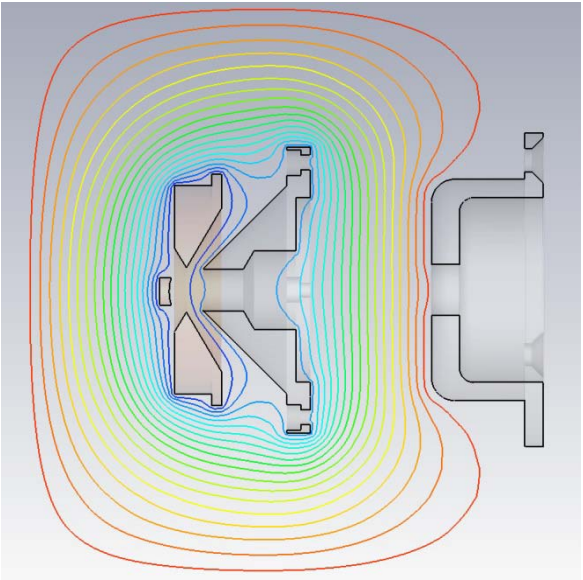

Fig.3. Electric potential lines of CYCLONE30 extraction system.

## 3 Ion beam extraction

A thorough experimental analysis of extraction optics for a single circular aperture is given by Coupland [3]. For uniform emission from an infinite plane, the Child–Langmuir law [4] gives the current density, J, of singly ionized hydrogen as

$$J = \frac{1.74}{d^2} V^{3/2} \text{ mA mm}^{-2} \quad (1)$$

where V is the applied extraction potential in kilovolt and d is the separation between the aperture and the extraction electrode in millimeters. For emission from a concave surface with radius of curvature $R_M$, for small values of $d/R_M$ it can be shown that Eq. (1) is multiplied by a factor

$$[1-1.6 d/R_M] \quad (2)$$

which is smaller than unity. The total beam current is found by multiplying the current density by the area of the emission aperture. For the cylindrical aperture of the ion source, the current is:

$$I = J\pi r^2 = 1.74 \pi V^{3/2} S^2 [1-1.6 d/R_M] \quad (3)$$

where $S = r/d$ is the aspect ratio, r is the radius of the hole in the plasma electrode, and $F = \pi r^2$ is the emitting area.

The perveance, P, of an ion beam is defined as

$$P = \frac{I}{V^{3/2}} = \left[1 - 1.6 \frac{d}{R_M}\right] P_0 \quad \text{mA kV-3/2} \quad (4)$$

which is a function only of the geometry of the system.

The divergence angle ω at the exit of the extraction system is caused by the shape of the plasma meniscus, by the defocusing forces of the second aperture as described above, by the temperature of the ions as we will see below, and by repulsive forces of the particles on themselves. The divergence angle has been calculated by Coupland [4],

$$\omega = 0.29\ S\ (1-2.14 P/P_0) \quad (5)$$

where $S=r/d$ as above, and P is the perveance in the extraction gap, $P=I/V^{3/2}$, and P0 is the Child-Langmuir space-charge limited perveance for the one dimensional diode of length d with no electrons,

$$P0 = [(4/9)\ \pi]\ (r^2/d^2)\ \varepsilon_0 (2e/M)^{1/2}, \quad (6)$$

where $\varepsilon_0$ is the permittivity of the vacuum.
Equation (5) predicts that the divergence can be reduced to zero at a perveance equal to 0.47 $P_0$. In practice the divergence does not decrease to zero and the perveance at the minimum divergence is less than 0.47 $P_0$.

## 4 Numerical simulation with CST code

Beam simulation codes have been valuable tools in understanding and designing ion source extraction and beam transport systems. There are quite a number of different codes developed and used within this community.

The CST code [6], a program developed for simulating ion beam optics in a certain accelerator configuration, has been widely used in designing the CYCLONE30 ion source accelerator.
CST PARTICLE STUDIO (CST PS) is a specialist tool for the fast and accurate analysis of charged particle dynamics in 3D electromagnetic fields. Powerful and versatile, it is suitable for tasks ranging from designing magnetrons and tuning electron tubes to modeling particle sources and accelerator components.

The particle tracking solver can model the behavior of particles through static fields, and with the gun iteration, space charge limited emission. The particle-in-cell (PIC) solver, which works in the time domain, can perform a fully consistent simulation of particles and electromagnetic fields. For relativistic applications, the wakefield solver can calculate how the fields generated by particles traveling at (or close to) the speed of light interact with the structure around them.

CST PS is integrated with the multi-purpose 3D EM modules of CST STUDIO SUITE, such as the CST EM STUDIO electro- and magnetostatic solvers and the CST MICROWAVE STUDIO eigenmode solver. It is fully embedded in the CST STUDIO SUITE design environment, thus benefitting from its intuitive modeling capabilities and powerful import interfaces. CST PS is based on the knowledge, research and development that went into the algorithms used in the MAFIA-4 simulation package. The powerful PIC solver can also make use of GPU computing, offering significant performance enhancements on compatible hardware.

Simulation of extraction systems using this simulation code is presented. Since there are many variables affecting the beam optics in a triode electrode accelerator, it is reasonable that the simulation study is concentrated on parameters related to the G1 electrode which most strongly affect the beam current and divergence. One of the easiest ways to augment the extracted beam current is increasing the transparency of the accelerator by enlarging the aperture diameters. However, with this solution, there is a concern about deterioration in the beam optics. For that reason we don't consider this case. The ideas applied to the accelerator design are as follows: shaping the electrode angle, reducing the electrode (G1) thickness, decreasing the G1-G2 electrodes gap, and adjusting the electrode G2 voltage.

### 4.1 Plasma electrode inclination angle

It has been accepted that shaping the G1 electrode, in general, is helpful to extract a beam at a low divergence angle. A well-known study by Pierce [7] found that a zero divergence electron beam can be extracted from either a slit or a cylindrically symmetric extractor if the plasma and extraction electrodes are shaped so as to match a Laplace solution outside to a Poisson solution inside the beam. The match demands the plasma electrode to have an inclination of 22.5°. H⁻ ion beams, however, are produced and subsequently behave very differently than electron and even proton beams, so it is not so clear what the optimum plasma electrode angle is. Various angles including 22.5°, 30°, and 45° have been used here [8,9]. Table 1 shows the extraction system simulation results with varying plasma electrode inclinations. A beam divergence has the lowest value when the shaping angle of the backside edge of an aperture is about 67°. The perveance of the present geometry is 4.94E-10 A/V$^{3/2}$ and beam divergence angle is about 3.14° which is quite a large value.

Table1: extraction system simulation results with varying plasma electrode inclinations.

| Angle of inclination (degree) | Beam divergence angle (degree) | Perveance (A/V3/2) |
|---|---|---|
| 30° | 3.14° | 4.94E-10 |
| 22.5° | 2.2° | 5.11E-10 |
| 45° | 3.32° | 4.57E-10 |

### 4.2 Plasma electrode thickness

The simulation results for the effect of the G1 thickness on the perveance and beam divergence is shown in Fig.4, where the divergence angle is given in an rms value. The optimum perveance increases with a decrease of the G1 thickness. A reduction of the G1 thickness also results in lowering the divergence angle. Thickness of 7mm is the optimum point which the maximum perveance and minimum divergence is observed.

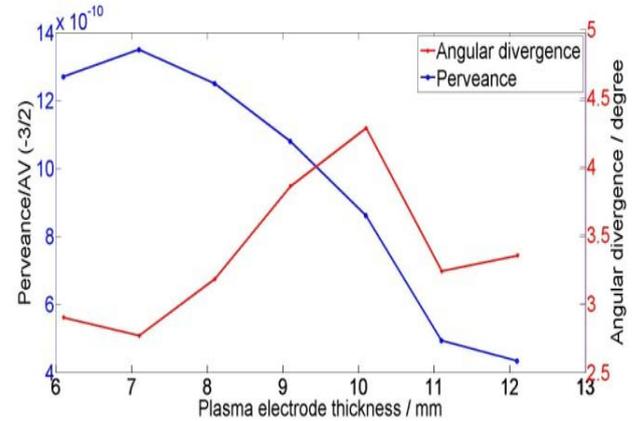

Fig.4. Perveance and angular divergence versus plasma electrode thickness variation.

### 4.3 The first gap distance

In Fig. 5, any reduction of the first gap raises the optimum perveance value and decreases the divergence angle of the beam. These tendencies can be explained reasonably in terms of familiar Child–Langmuir's law. Here we denote the first distance as $D_{12}$ and 2.5 mm gap distance is preferred.

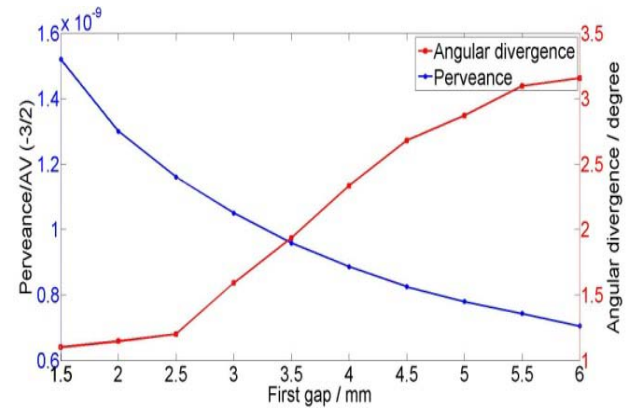

Fig.5. Perveance and divergence angle versus first gap width.

### 4.4 Second Electrode Voltage

Fig. 6 summarizes the simulation results for the effect of the variation in the voltage of the G2 electrode, V2. Increasing the voltage results in higher perveance but deteriorates the divergence, which points out that there is an optimum point for the second electrode voltage compromising contradictory trends of two parameters. The potential of -24000V is chosen for the second electrode to have a reasonable angular divergence and perveance, i.e, a beam with high perveance and not too bad divergence.

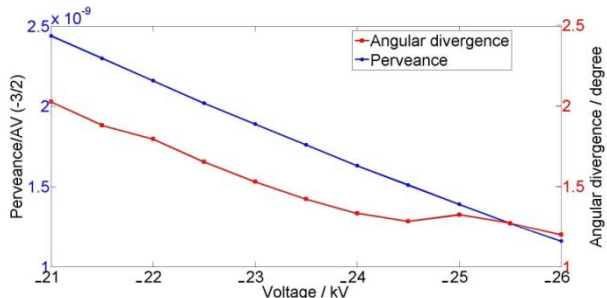

Fig.6. Perveance and divergence angle versus second electrode voltage variation.

## 5 Discussion

Fortunately, we could make a desirable accelerator design that satisfies contradictory goals: high perveance and good optics. From the above simulations, an idea has been given in regard to reforming the accelerator configuration, which includes both sliming the G1 electrode to a 7 mm thickness and shortening the second gap to 2.5 mm. Fig. 7 is the calculated profile of the extracted ion beam in the upgrade accelerator structure, which shows acceptable beam optics properties of enlarged electrode holes. The perveance of present geometry is 1.63E-009 A/V^(3/2) and its divergence angle is 1.3° which considerably has been improved.

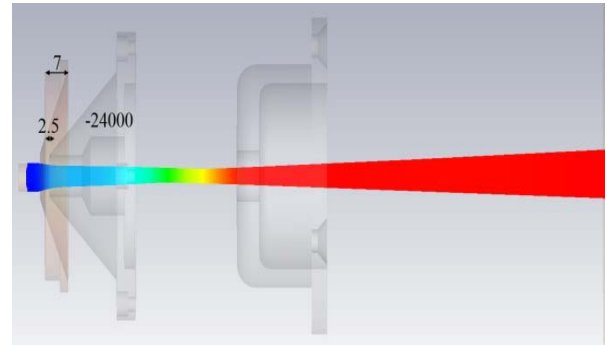

Fig.7. Simulation runs for the upgrade accelerator structure.

## 6 Conclusion

In this paper, the ion beam trajectories for different parameters of the accelerator structure were simulated and optimized. CST simulation provides some conclusions: first, shaping the plasma electrode angle, second, reducing the thickness of the plasma electrode is helpful for a high perveance beam extraction to maintain a low divergence. Third, by decreasing the first gap both a high optimum perveance and improved beam optics can be attained. Forth, by optimizing the G2 voltage has both an advantage of high optimum perveance and a disadvantage of bad optics. From the above simulation results, there is now an idea for reforming the accelerator configuration, which includes sliming the G1 electrode to a 7 mm thickness, shortening the first gap to 2.5 mm to obtain a good beam current and increasing the G2 voltage.